\documentclass[12pt,preprint]{aastex}
\usepackage{emulateapj5,graphicx,times}

\shorttitle{Halpha emitting regions in MS IV}
\shortauthors{Yagi, Komiyama \& Yoshida}
\begin{document}

\title{Candidates of Halpha emitting regions in Magellanic Stream IV cloud} 

\author{Masafumi Yagi\altaffilmark{1,2},
Yutaka Komiyama\altaffilmark{2},
Michitoshi Yoshida\altaffilmark{3}}

\altaffiltext{1}{email:YAGI.Masafumi@nao.ac.jp}
\altaffiltext{2}{
Optical and Infrared Astronomy Division,
National Astronomical Observatory of Japan,
2-21-1, Osawa, Mitaka, Tokyo, 181-8588, Japan}
\altaffiltext{3}{
Hiroshima Astrophysical Science Center, Hiroshima University,
1-3-1, Kagamiyama, Higashi-Hiroshima, Hiroshima, 739-8526, Japan
}

\def\Ha{H$\alpha $}
\def\Hb{H$\beta $}
\def\HI{H\,{\sc i}}
\def\HII{H\,{\sc ii}}
\def\NII{[N\,{\sc ii}]}
\def\OI{[O\,{\sc ii}]}
\def\SII{[S\,{\sc ii}]}


\begin{abstract}
From \Ha\ narrow band observations, we identified three 
\Ha\ emitting regions in the direction of Magellanic Stream IV (MS IV). 
They consist of three parallel filaments with 
2 arcmin width and 6 -- 30 arcmin length at 12 arcmin intervals.
The mean surface brightness of them is $\sim 2 \times 10^{-18}$
erg s$^{-1}$ cm$^{-2}$ arcsec$^{-2}$.
Because of their low surface brightness,
the regions were not detected in previous \Ha\ surveys.
In \HI\ map, the position of the filaments overlap MS, suggesting that 
they are parts of MS, but there also exists a local \HI\ structure.
If the filaments associate with MS, the sizes are 30 pc $\times$ 
100 -- 500 pc.
The filaments lie at the leading edge of a downstream cloud,
which supports a shock heating and its propagation (shock cascade) model
for the ionizing source.
If they are local objects, on the other hand, Fossil Str\"omgren Trails
of more than two stars is a possible interpretation,
and the sizes would be 0.1 pc $\times$ 0.3 -- 1.5 pc at 180 pc distance.
The positional information of the \Ha\ filaments presented in this letter 
enables us future spectroscopic observations to clarify their nature.
\end{abstract}



\section{Introduction}
Magellanic Stream (MS hereafter) was found as a long filament of \HI\ gas 
around our Galaxy \citep{Mathewson1974}.
MS is thought to be stripped gas from 
Small Magellanic Cloud and/or Large Magellanic Cloud.
It has been revealed that MS consists of multi-phase gas
components; molecular gas, \HI\ gas, \HII\ gas, and highly ionized
plasma \citep[and references therein]{Fox2010}.
MS is the only clear example of a gaseous halo stream 
in the close proximity of Milky Way \citep{Stanimirovic2008}.
It gives us important clues to the physical interaction between
Galaxy and its satellites in detail.

The \Ha\ emission from MS has been first reported by 
\citet[hereafter WW96]{Weiner1996}.
\citet{Putman2003b} have found several other \Ha\ emitting regions.
The ionizing source of these regions is still a mystery because 
no ionizing star is observed in the stream.  
There are several models to explain the origin of the MS \Ha\ clouds. 
WW96 argued that the emissions are explained by ram pressure heating.
However, this model required much higher density of halo gas than 
WW96 estimated, because the distance to MS was found to be larger than
expected \citep{Weiner2001}. \citet{BlandHawthorn1999}
presented another model that ionizing photons escaping from Galactic
disk heat up MS. This model predicted that the \Ha\ emission
would be smoothly distributed. However it is inconsistent with the 
fact that the contrast of the \Ha\ emission is as large as a factor of
30 \citep{Weiner2001}.
\citet[hereafter B07]{BlandHawthorn2007}
presented yet another model incorporating a shock heating and its
propagation (shock cascade). They predicted high \Ha/\Hb\ ratio, and
strong \OI, \SII\ emissions, which have not been examined yet.
Recently, \citet{Fox2010} investigated the absorption features of 
active galactic nuclei behind MS in order to study the 
ionization structure of the MS \Ha\ clouds. 
They found that the ionization structure cannot be reproduced by
an evaporative interaction with the Galactic corona plus photoionization.

In order to study the ionization mechanism and kinematics of the 
MS \Ha\ clouds, detailed spectroscopic follow-up observations are
required.
However, positional information necessary to perform a 
medium to high dispersion spectroscopy has been so far very limited.
The exact position and detailed morphology of the \Ha\ emitting region
found by WW96 was not clear, because 
they used
a Fabry-P\'erot interferometer and the spatial structure inside the
Fabry-P\'erot beam of 7 arcmin diameter was averaged. 

There are several other \Ha\ surveys which overlaps the MS fields.
Southern \Ha\ Sky Survey Atlas\footnote{http://amundsen.swarthmore.edu/} 
 \citep[SHASSA;][]{Gaustad2001} presents \Ha\ map of 0.8 arcmin pixel.
They used a 3.2 nm width filter to detect \Ha\ emissions.
Wisconsin \Ha\ mapper\footnote{http://www.astro.wisc.edu/wham/}
\citep[WHAM;][]{Raynolds1998,Haffner2003} 
provides another \Ha\ map with high velocity resolution in 
a velocity range from $-80$ km s$^{-1}$ to $+80$ km s$^{-1}$ with
respect to the local standard of rest (LSR)
with 1 degree beam using a Fabry-P\'erot spectrometer.
The spatial resolutions of those works are still not enough for a 
slit spectroscopy.

We performed imaging observations with higher spatial resolution
at the northern part of MS IV where WW96 detected the \Ha\ emissions.
In this letter, we report the positions and the shapes
of the \Ha\ emitting regions we detected.

\section{Observation}

We observed two WW96 fields, MS IV-C (23:44:34.8,-12:23:20,J2000.0) 
and MS IV-D (23:40:49.0,-10:45:32,J2000.0) with 2kCCD camera 
\citep{2kCCD} mounted on Kiso 105 cm Schmidt camera at the Kiso Observatory
in September 2011.
The pixel scale is 
1.5 arcsec pixel$^{-1}$,
and the field of view (FoV) is 51.2 arcmin square.
We used R-band and three narrow band filters (N657, Ha6577, and Ha6417).
The filter characteristics are shown in Table 1.
The variation of the wavelength coverage
is no worse than 3.6\AA\ across the filter in Ha6577 and Ha6417.
For N657, the transmission curve at the corner 
is shifted by 10\AA\ to the red from that at the center.
We took the exposures with $\sim$ 7.5 arcmin dithering.
The typical seeing size was 4.1 arcsec.

The data reduction was performed 
using the mosaic CCD data reduction packages \citep{Yagi2002}.
Overscan subtraction, bias pattern subtraction, 
flat-fielding with domeflat were performed in a standard manner.
The pixel scale of the images are slightly different among filters.
The difference was corrected before coadding
using the results by SMOKA team
(S. Ichikawa, private communication)
\footnote{http://smoka.nao.ac.jp/about/KCD\_astrometric\_calib.jsp}.
The sky background was subtracted in each frame before coadding with 
background mesh size of 1024 pixels (25.6 arcmin).
Therefore, structures larger than 25 arcmin square 
would have been removed.
After the coadding, stellar objects were detected by
SExtractor \citep{Bertin1996}, and 
the astrometric calibration was performed  
using wcstools \citep{Mink2002} referring to USNO-B1 catalog \citep{Monet2003}.


We calibrated the flux using the photometry of stars 
by Sloan digital sky survey (SDSS) 
data release 7 \citep{DR7} in the northern half of 
the MS IV-D field.
The color conversion from SDSS r and i to Kiso 2kCCD filters
are constructed by fitting a quadratic function
to the photometry of model spectral energy distributions (SEDs)
by Bruzual-Pearson-Gunn-Stryker atlas
\footnote{ftp://ftp.stsci.edu/cdbs/grid/bpgs/}
and the filter responses, as the method in the appendix of \citet{Yagi2010}. 
The best-fit functions are
\begin{equation}
Kiso=r+c_0+c_1(r-i)+c_2(r-i)^2,\\
\end{equation}
where $(c_0,c_1,c_2)$=
(0.003,0.267,0.052),
(-0.075,0.773,-0.292),
(-0.034,0.653,-0.210),
and 
(0.014,0.262,0) for R, N657, Ha6577, and Ha6417, respectively,
in AB-magnitude for $-0.1<r-i<1.0$ stars.
The dispersion of our photometry around the fit is 0.06 -- 0.1 mag in rms 
converted from the median of absolute deviation in each band.
As MS IV-C is not covered by SDSS, 
we roughly estimated the zero point using 
R-band ($\rm R_F$-band) photometry of USNO-B1.0 \citep{Monet2003} 
in Naval Observatory Merged Astrometric Dataset
\citep{Zacharias2005}. 
The error is estimated to be $\sim$ 0.15 mag.
Galactic extinction is about $A_R=0.08-0.09$ in MS IV-C and MS IV-D
from \citet{Schlegel1998} via the extinction calculator in
NASA/IPAC extragalactic database (NED)\footnote{http://ned.ipac.caltech.edu/forms/calculator.html}.
We adopted 0.09 mag for all of the 4 bands.
The limiting surface brightness (SB) of the final images of MS IV-D
in 15 arcsec (10 pix) square are presented in Table 1.

\section{Detection of three \Ha\ emitting regions}

We performed eye inspections to detect possible \Ha\ emitting regions. 
In the MS IV-C field,
we could not find any sign of \Ha\ excess brighter than 800 mR
\footnote{1 Rayleigh (1R) is 10$^{10}$ 
photons m$^{-2}$ s$^{-1}$ str$^{-1}$, which corresponds to 
$5.66 \times 10^{-18} ~{\rm erg} ~{\rm s}^{-1} ~{\rm cm}^{-2} 
~{\rm arcsec}^{-2}$ at 6563\AA, and 1 R=1000 mR.}
with size between 15 arcsec and 25 arcmin scale
in the 54 $\times$ 54 arcmin square field.
The conversion from SB of N657-R to total \Ha+\NII\ brightness
requires the redshift and the line full width at half maximum (FWHM) 
information, which are unavailable yet.
In this study, we therefore assume that the line is located at the
wavelength of the maximum transmittance of each filter,
and that the line width is much narrower than the filter FWHM.
This gives the lower limit of the \Ha+\NII\ brightness.

In the MS IV-D field, we detected 
regions with N657 excess relative to R-band.
Fig. 1a and 1b show the R-band subtracted N657 image (N657-R image) 
and the N657 image of the MS IV-D field, respectively.
The images were binned by 10$\times$10 pixels (15 arcsec square)
to increase signal to noise ratio (S/N). 
Three filamentary \Ha\ regions are seen in the lower half
region of Fig. 1a and 1b.
We call the \Ha\ regions as
D1, D2, and D3 from the east to the west, hereafter.
Except the three, we could not detect \Ha\ emitting structures
brighter than 240 mR (1$\sigma$ of the N657-R image)
with size between 15 arcsec and 25 arcmin scale
in 74 $\times$ 60 arcmin field of MV IV-D.
Fig. 1c shows the effective exposure time map of N657-R and 
the position of the \Ha\ regions.
Because of the bad weather, the southwest part has low S/N.
It should also be noted that a bright star at the southwest of D2 
made it difficult to measure the emission around it.

We found D1 and part of D3 in the Ha6577 image
with SB of 26.3 mag arcsec$^{-2}$.
The three \Ha\ regions are fainter than 27 mag arcsec$^{-2}$ in R-band.
In the Ha6417 image, neither D1 nor D2 were detected,
and D3 was located outside of the FoV.
D1 was also found in the N657-Ha6417 image, and marginally recognized in 
the Ha6577-Ha6417 image.
In the Ha6577-R image, 
D1 and D3 were detected, and D2 was marginally detected. 
In the N657-Ha6577 image, D1 was marginally positive (N657 bright) with low S/N,
and D2 and D3 were not detected. 
In summary, the excess was found in N657-R, N657-Ha6417, and Ha6577-R,
and marginally in N657-Ha6577. The detection in several combination of 
bands suggests that the features are real emissions.

We investigated the possibility of artifacts.
If the filaments were fringes made by sky lines,
they should also appear in other fields, 
but we could not detect such features in other fields (e.g., MS IV-C).
We also checked whether the regions were seen in coadded images of each day,
and confirmed that they were not moving objects nor transient phenomena.
The positions and the shapes remained the same during the observation.
When we observe other brighter stars in another field,
we found some ghosts of the optics appear. The ghosts moved
when we shifted the FoV,
and they appeared at different position in Ha6577 and N657. 
On the contrary, D1--D3 did not move
and stayed at the same position in the sky in both filters.
Moreover, we made additional observations in November 2011 for the
filaments with Ha6577. At that time, we shifted the field to avoid
the bright star at south-east. 
The filaments D1 and D3 were still detected at the same position.
D1--D3 are therefore not ghosts.
We conclude that the three regions are genuine \Ha\ emitting regions.

The flux weighted center of the regions are shown in Table 2.
The positions are 20 -- 50 arcmin away from the 
aperture of MV IV-D by WW96 (Fig. 1c).

We defined the boundary of the filaments 
for a quantitative analysis as follows.
First, we masked bright objects in R-band,
because a residual of subtraction may remain in 
the N657-R image at the position. 
Next, we applied a 7$\times$7 pix (10.5 arcsec square) 
median filtering to gain S/N. 
Then the R-band image is subtracted from the N657 image, and binned 
into 10$\times$10 pix (15 arcsec square).
In the image, we defined the boundary as 
the isophote of 220mR, which corresponds to 1-$\sigma$.
Above the isophote the regions have $\sim$2 arcmin wide
and 16, 6, and 29 arcmin long for D1, D2, and D3, respectively.
The interval between D1 and D2 are about 12 arcmin, 
and that between D2 and D3 is also $\sim$12 arcmin.
If the regions lie at MS distance \citep[55 kpc;][]{Putman2003a}, 
the width of the three \Ha\ regions 
is 30 pc, intervals between them are 200 pc each, 
and the length are 100 -- 500 pc.

For photometric analysis, we used the isophote as 
the boundary and extracted the regions before median filtering.
The SB of the \Ha\ regions in the N657-R image are 370 mR on average
in 15 arcsec square, which corresponds to \Ha+\NII\ brightness of
$\sim 2.1 \times 10^{-18} 
~{\rm erg} ~{\rm s}^{-1} ~{\rm cm}^{-2} ~{\rm arcsec}^{-2}$.
The SB gradient is seen in D1 and D3. 
The north-west edge is the faintest in SB in D3, 
and the gradient is about 20 mR arcmin$^{-1}$.
The gradient in D1 has an opposite direction;
the SB is the faintest at the southeast edge,
rises by 60 mR arcmin$^{-1}$
for 10 arcmin, then gradually get fainter toward north west.
The gradient is not seen in D2.

\section{Comparison with \Ha\ surveys} 

Our observed field was covered by two \Ha\ surveys.
One is SHASSA \citep{Gaustad2001}.
The \Ha\ bandwidth (3.2nm) of SHASSA is comparable to our N657.
\citet{Finkbeiner2003} removed stellar contamination from 
the SHASSA data and calibrated the flux\footnote{http://skymaps.info/}.
There are several bright parts in SHASSA data which overlap D1--D3.
However, when we checked the original data of SHASSA, they are
not permanent but variable among exposures. The bright parts would
be artifacts.

The other survey is WHAM, which covers a velocity range of 
$\sim \pm 80$km s$^{-1}$ with respect to LSR.
One of WHAM beam, $(l,b) =$(73.377, -67.054) is just on D1
\citep{Haffner2003}.
If we average the \Ha+\NII\ flux of the regions D1--D3 in 
1 degree aperture of WHAM, it would be observed
as $\sim$ 5 mR excess, which is too faint to be found in WHAM.

\section{Comparison with \HI\ map}

We overlap the position of the \Ha\ regions 
on the \HI\ contour using Galactic All-Sky Survey (GASS) Second Data Release 
\citep{McClureGriffiths2009,Kalberla2010}\footnote{http://www.astro.uni-bonn.de/hisurvey/}
in Fig. 2.
The regions D1--D3 lie on an overlap of two \HI\ components at 
$v_{LSR}\sim-210$ km~s$^{-1}$ and $v_{LSR}\sim-30$ km~s$^{-1}$.
We thus coadd the GASS data over two velocity ranges;
$-170 - -250$ km~s$^{-1}$ as the high velocity component(HVC), and
$-15 - -63$ km~s$^{-1}$ as the low velocity component(LVC).

In HVC map (Fig 2a),
an \HI\ cloud with a horseshoe like morphology
is located at $(l, b) = $(73.5, -68) \citep{Bruns2005},
which is at the southeast of our observed field.
The \HI\ cloud is also recognized in the $-211.2$ km~s$^{-1}$ panel
of Fig. 6 of \citet{Putman2003a}.
They mentioned that filaments of MS
appear to merge at several points, and 
there are dense concentrations of gas at these points.
They listed the cloud at (l,b)=(74,-68) as 
an example of the merging points.
The regions D1--D3 are located on this cloud.
Spatial correspondence between D1--D3 and the
\HI\ cloud suggests that the \Ha\ regions are associated with MS,
and we regard them as candidates of \Ha\ emitting regions of MS. 
The \HI\ column density around the \Ha\ regions are 
$2 - 6 \times 10^{19}$ cm$^{-2}$ from the contour by \citet{Bruns2005}.

In LVC (Fig. 2b), there are two clouds around the edge of the filaments,
and it is also possible that D1--D3 are associated with LVC.
The \HI\ column density around the \Ha\ regions are 
$3 - 6 \times 10^{19}$ cm$^{-2}$.

\section{Column density of the ionized gas}

We roughly estimate the column density of the ionized gas.
As the typical flux is 370 mR including \NII,
we adopt 300 mR for \Ha\ flux in this section,
assuming \NII/\Ha\ $\sim 0.2$ 
\citep{Ly2007} for simplicity.
The value is a typical one for \HII\ regions,
and it would be largely different if the ionization process
is different, such as shocks. 
We assume that the \Ha\ emitting region 
is a homogeneously distributed, optically thin gas cloud, 
and is embedded in an isotropic radiation field.
For \Ha, 1 Rayleigh corresponds to an emission measure (EM)
of about 2 cm$^{-6}$ pc\citep{McCullough2001}.
Therefore the EM of D1--D3 is $\sim$ 0.6 cm$^{-6}$ pc.
The EM is written as 
\begin{equation}
EM=\langle n_e^2 \rangle L = 0.6 ~{\rm cm}^{-6} {\rm pc},
\end{equation}
where $n_e$ is the electron density, and 
$L$ is the mean length of line of sight.
The column density is then
\begin{equation}
\sqrt{\langle n_e^2 \rangle} L = 7.6 \times 10^{18}
\left(\frac{L}{10{\rm pc}}\right)^{1/2} {\rm cm}^{-2}.
\end{equation}
It is about 1/2$-$1/8 of \HI\ column density ($2 - 6\times 10^{19}$ cm$^{-2}$),
if $L=10$ pc. 

This rough calculation may overestimate the column density,
because, in the real, the \Ha\ surface brightness distribution
is not homogeneous.
In addition, the redshift and the degree of \NII\ contamination
are unknown.

\section{Comparison with model by B07}

Since WW96 discovered the \Ha\ emission from MS, 
several hypotheses have been proposed for the ionization
mechanism of the \Ha\ emitting gas.
A promising theory would be the shock-induced model presented by B07.
They suggested that the interaction between the fragmented
clouds formed in the upstream of MS and the tenuous gas
ablated from the upstream clouds would lead 
to shock ionization at the leading edges of the downstream clouds.

If the northwest cloud in Fig. 2a, which lies in the 
downstream of the 
main MS IV cloud at southeast, is following the main cloud, 
\Ha\ would be emitted at the leading edge of the northwest cloud
according to the B07 model.
On the contrary, D1--D3 appear to lie along the \HI\ stream 
between the northwest and the main clouds.
Their straight and almost parallel filamentary morphology 
do not resemble the \Ha\ distribution predicted by B07.
However, note that the widths of D1--D3, $\approx 30$ pc,
are smaller than the spatial resolution of the simulations
($\approx 40$ pc) of B07.
In addition, kpc scale structures of the \Ha\ emission may be lost 
in this study due to our sky subtraction method (see Section 2).
Thus, the spatial resolution mismatch between our observation and the B07
simulation may lead this apparent inconsistency in ionized gas
morphology. 
The \Ha\ regions we detected may be a part of 
dense ridges, which can be formed by Kelvin-Helmholtz instability
at the \HI\ cloud surface, of tenuously extended ionized clouds.  

\section{Possibility of Fossil Str\"omgren Trail}

The long and straight shape of D1--D3 resembles 
the \Ha\ filament reported by \citet[hereafter MB01]{McCullough2001}.
The MB01 cloud has a region of 20 arcsec wide and 1.2 degree long and 
forks into two segments with the widest separation of 5 arcmin.

MB01 presented four possible origins for such a cloud; 1) a jet, 2) 
a filamentary local nebula, 3) a shock ionized trail, and 
4) a Fossil Str\"omgren Trail (FST),
and they concluded their object would be FST.
If D1--D3 are local objects, the origin of D1--D3 would be one of the four.
Following the discussion by MB01, 
the faint SB rejects 1), and the straight shape and 
the intensity gradient along the shape reject 2), 
and 2 arcmin width rejects 3) as the D1--D3 origin.
And 4), FST by a single star, is difficult,
since the gradient has opposite direction in D1 and D3.
If D1--D3 are local objects, the possible origin is 
that they are FSTs of different stars.
All the three might be made by three different stars 
or by two stars if D2 is a branch of D1 or D3
like the Y-shaped filament found by MB01.

If D1--D3 are FSTs, we can estimate the distance using equation (5) of 
MB01. Putting I(\Ha)$_{\rm obs}$=300 mR, width $\gamma$=2 arcmin,
and the assumption by MB01 that projection factor $p$=1, 
temperature T=$10^4$K, and density n=1 cm$^{-3}$, 
we can estimate that the distance is
$\sim$180 pc. The physical size is 0.1pc wide and 0.3 -- 1.5 pc long.

\section{Summary}

We detected three filamentary 
\Ha\ emitting regions in the MS IV-D field.
Their sizes are 2 arcmin wide and 6 -- 30 arcmin long,
and their surface brightness 
of \Ha+\NII\ are 370 mR on average in 15 arcsec bin.
They are candidates of \Ha\ emitting regions 
associating with MS IV, although we cannot deny the possibility 
that they are local objects.
If they are at MS distance, the size is 30 pc $\times$ 100 -- 500 pc.
If they are local objects, 
Fossil Str\"omgren Trails \citep{McCullough2001} 
created by more than two stars is a possible interpretation.
Future spectroscopic follow-up will reveal whether they are a part of MS.

\acknowledgments

We thank the anonymous referee for 
valuable comments and suggestions.
We thank staff at Kiso Observatory for their kind supports.
We thank Christian Br\"uens, Juergen Kerp,
and Peter M. W. Kalberla for important comments and 
suggestions.
We thank Shin-Ichi Ichikawa and Satoshi Kawanomoto
for information about data reduction. 
This study was partly supported by
KAKENHI 21540247 and 23244030.
This work has made use of the 
SDSS 
\footnote{http://cas.sdss.org/}, 
NED 
\footnote{http://nedwww.ipac.caltech.edu/}, 
SMOKA archive 
\footnote{http://smoka.nao.ac.jp/}
and the computers 
at ADC NAOJ.


\onecolumn

\begin{figure}
\includegraphics[scale=0.18,bb=14 14 962 1101]{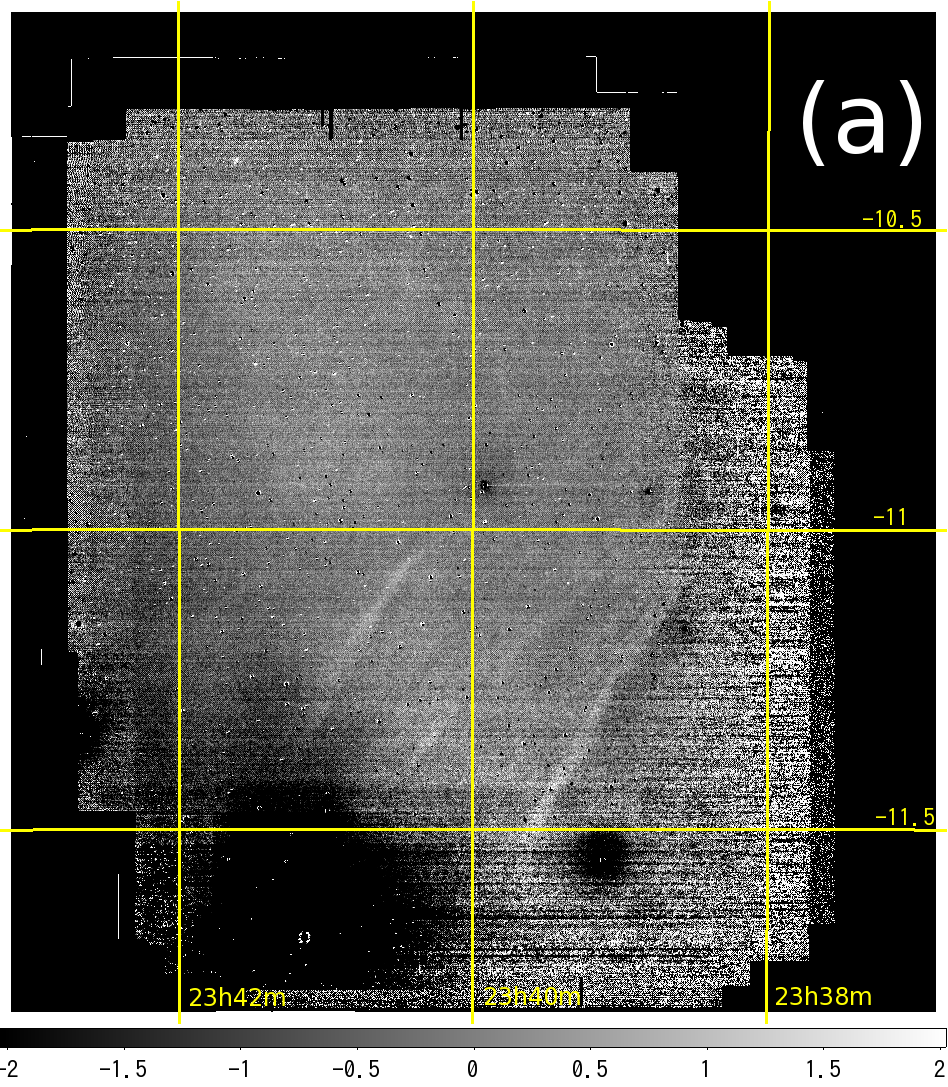}
\includegraphics[scale=0.18,bb=14 14 962 1101]{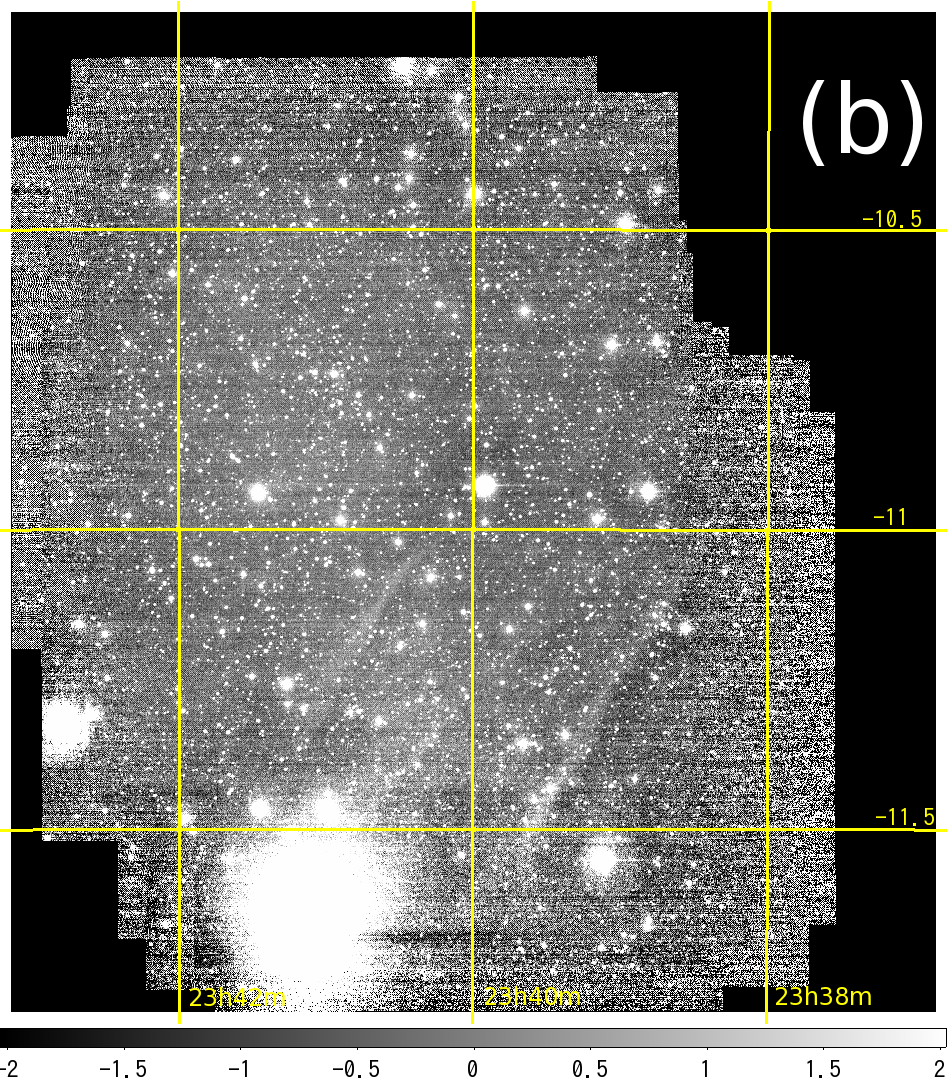}
\includegraphics[scale=0.18,bb=14 14 962 1101]{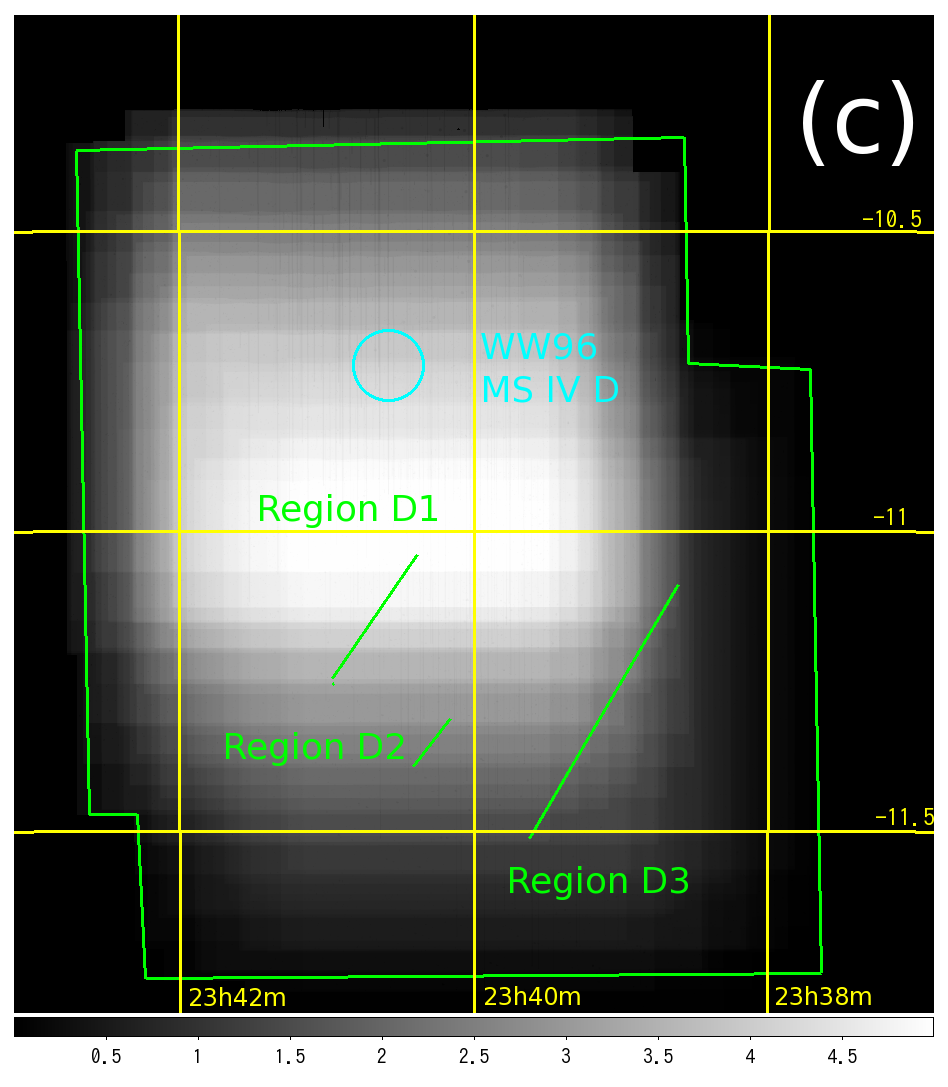}
\caption{Coadded images of the MS IV-D field: N657-R (a) and N657 (b).
North is up, and east to the left,
and the coordinates are J2000.0.
The image size is 87.5 $\times$ 100 arcmin.
The scale of colorbar is in units of Rayleighs.
Effective exposure time map of N657-R is shown as (c).
The colorbar is in units of hours.
The aperture of WW96 (cyan) and 
the position of the \Ha\ emitting regions (green) are overlaid.
}
\end{figure}

\begin{figure}
\includegraphics[scale=0.2,bb=14 14 1026 832]{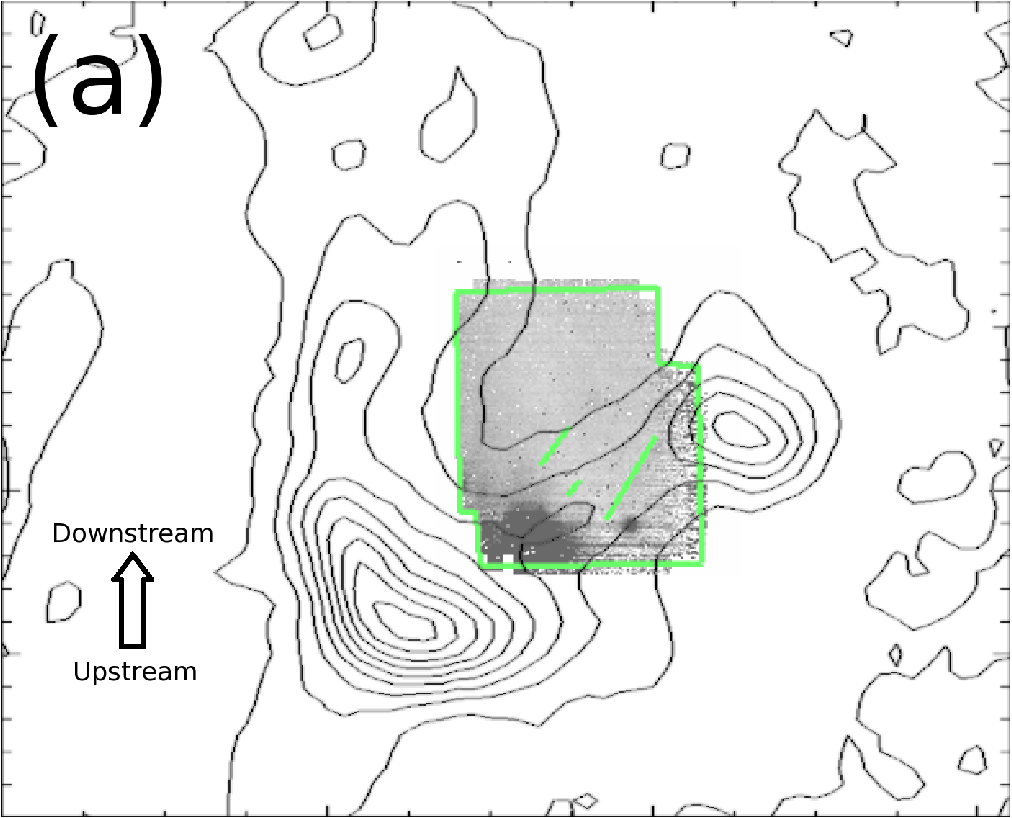}
\includegraphics[scale=0.2,bb=14 14 1026 832]{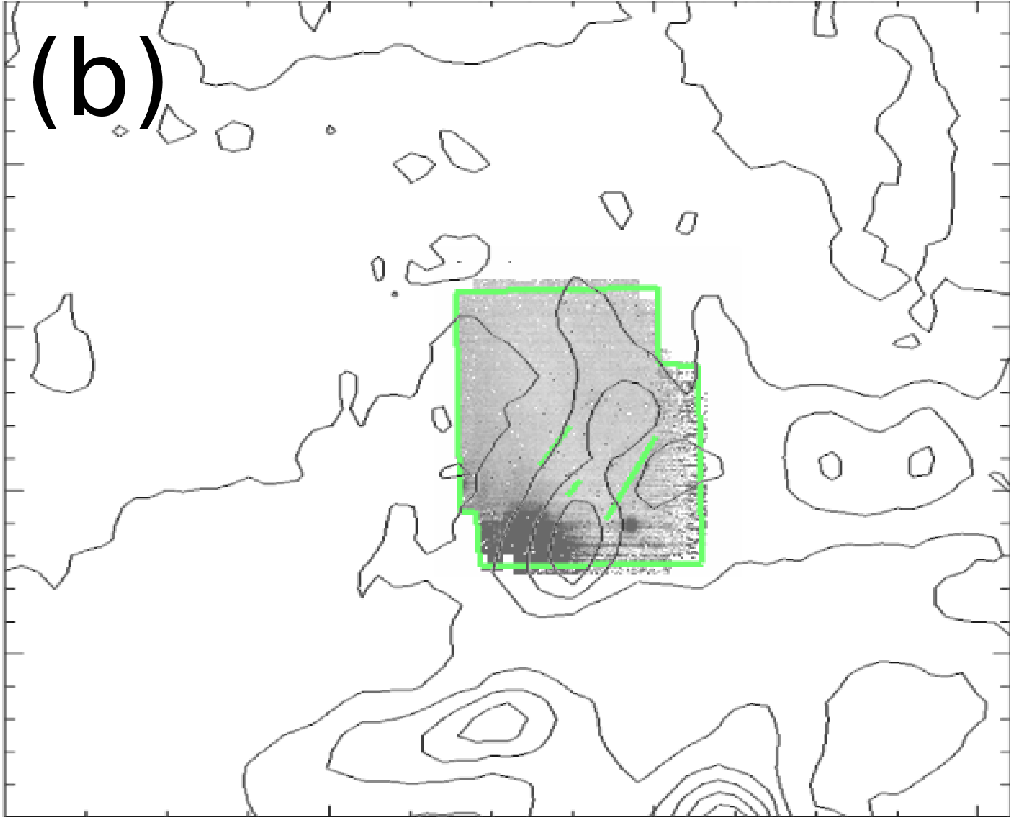}
\caption{Overlay of N657-R map in this study with D1--D3 mark on
the \HI\ contour. The size is 5 $\times$ 4 degree, north is up
and east to the left.
The \HI\ contour is created from GASS Second Data Release
\citep{McClureGriffiths2009,Kalberla2010}.
(a) The velocity of $-170<v<-250$ km s$^{-1}$ are coadded.
The contour corresponds to 2$\times 10^{19}$ cm$^{-2}$ interval
from 2$\times 10^{19}$ cm$^{-2}$.
(b) The velocity of $-63<v<-15$ km s$^{-1}$ are coadded.
The contour corresponds to 1$\times 10^{19}$ cm$^{-2}$ interval
from 3$\times 10^{19}$ cm$^{-2}$.
}
\end{figure}

\begin{table}
\begin{tabular}{|l|c|c|c|l|}
\hline
filter & center & FWHM & v(\Ha) & 1$\sigma$ SB \\
\hline
R      & 6475\AA & 1450\AA & --- &27.4 mag arcsec$^{-2}$\\
N657   & 6577\AA & 40\AA & -265..+1563 km s$^{-1}$& 27.3 mag arcsec$^{-2}=210 mR$\\
Ha6577 & 6577\AA & 85\AA & -1294..+2592 km s$^{-1}$& 27.2 mag arcsec$^{-2}=500 mR$\\
Ha6417 & 6417\AA & 85\AA & -8608..-4722 km s$^{-1}$& 26.8 mag arcsec$^{-2}$\\
\hline
\end{tabular}
\caption{Used Filters, and 1-$\sigma$ surface brightness(SB)
of the MS IV-D field. 
The column v(\Ha) shows the recession velocity coverage for \Ha.
The 1-$\sigma$ SB is estimated around the 
highest flux part in Fig. 1c, and measured in 15 arcsec square.
For N657 and Ha6577, the SB is also given in units of milli-Rayleigh(mR).
}
\end{table}

\begin{table}
\begin{tabular}{|l|l|l|l|l|}
\hline
name &R.A.(J2000)&Dec.(J2000)& $l$ & $b$\\
\hline
MS IV-D1 & 23:40:43 & -11:09:06 & 73.3891 & -66.9449 \\
MS IV-D2 & 23:40:13 & -11:19:40 & 72.8502 & -66.9864 \\
MS IV-D3 & 23:39:17 & -11:21:00 & 72.3947 & -66.8430 \\
\hline
\end{tabular}
\caption{The flux weighted center of the \Ha\ emitting regions}
\end{table}

\begin{thebibliography}{}

\bibitem[Abazajian et al.(2009)]{DR7}
Abazajian, K. N. et al. 2009, \apjs, 182, 543

\bibitem[Bertin \& Arnouts(1996)]{Bertin1996}
Bertin, E., Arnouts, S. 1996, \aaps, 117, 393

\bibitem[Bland-Hawthorn \& Maloney(1999)]{BlandHawthorn1999}
Bland-Hawthorn, J., Maloney, P.R. 1999, \apjl, 510, L33

\bibitem[Bland-Hawthorn et al.(2007)]{BlandHawthorn2007}
Bland-Hawthorn, J., et al. 2007, \apjl, 670, L109(B07)

\bibitem[Br\"uns et al.(2005)]{Bruns2005}
Br\"uns C. et al. 2005, \aap, 432, 45 

\bibitem[Fox et al.(2010)]{Fox2010}
Fox, A.J., Wakker, B.P., Smoker, J.V.,
Richter, P., Savage, B.D, Sembach, K.R. 2010, \apj, 718, 1046 

\bibitem[Finkbeiner(2003)]{Finkbeiner2003}
Finkbeiner D. P. 2003, \apjs, 146, 407

\bibitem[Gaustad et al.(2001)]{Gaustad2001}
Gaustad, et al. 2001, \pasp, 113, 1326 

\bibitem[Haffner et al.(2003)]{Haffner2003}
Haffner, L.M., Reynolds, R.J., Tufte, S.L.,
Madsen, G.J., Jaehnig, K.P., Percival, J.W. 2003, \apjs, 149, 405

\bibitem[Itoh et al.(2001)]{2kCCD}
Itoh, N. et al. 2001, Publ. Natl. Astron. Obs. Japan, 6, 41

\bibitem[Kalberla et al.(2010)]{Kalberla2010} 
Kalberla, P.M.W., et al. 2010, \aap, 512, A17

\bibitem[Ly et al.(2007)]{Ly2007}
Ly, C. et al. 2007 \apj, 657, 738

\bibitem[Mathewson et al.(1974)]{Mathewson1974}
Mathewson, D. S., et al. 1974, \apj, 190, 291 

\bibitem[McCullough \& Benjamin(2001)]{McCullough2001}
McCullough, P.R., Benjamin, R.A. 2001, \aj, 122, 1500

\bibitem[McClure-Griffiths et al.(2009)]{McClureGriffiths2009}
McClure-Griffiths, N. M., et al. 2009, \apjs, 181, 398

\bibitem[Mink(2002)]{Mink2002}
Mink, D.J. 2002, in ASP Conf. Proc. 281. ADASS XI, 
ed. D.A. Bohlender, D. Durand, \& T.H. Handley (San Francisco: ASP), 169 

\bibitem[Monet et al.(2003)]{Monet2003} 
Monet, D.G. et al. 2003, \aj, 125, 984

\bibitem[Putman et al.(2003a)]{Putman2003a}
Putman, M.E., Staveley-Smith, L., Freeman, K.C., 
Gibson, B.K., Barnes, D.G. 2003a, \apj, 586, 170

\bibitem[Putman et al.(2003b)]{Putman2003b}
Putman, M.E., Bland-Hawthorn, J., Veilleux, S.,
Gibson, B.K., Freeman, K.C., Maloney, P.R. 2003b, \apj, 597, 948 

\bibitem[Raynolds, et al.(1998)]{Raynolds1998}
Raynolds, et al. 1998, \pasa, 15, 14

\bibitem[Schlegel et al.(1998)Schlegel, Finkbeiner, \& Davis]{Schlegel1998}
Schlegel, D. J., Finkbeiner, D. P. \& Davis, M., 1998, \apj, 500, 525

\bibitem[Stanimirorovi\'c et al.(2008)]{Stanimirovic2008}
Stanimirorovi\'c et al. 2008, \apj, 680, 276 

\bibitem[Weiner \& Williams(1996)]{Weiner1996}
Weiner, B.J., Williams, T.B. 1996, \aj, 111, 1156 (WW96)

\bibitem[Weiner et al.(2001)]{Weiner2001}
Weiner, B.J.,  Vogel, S.N., Williams, T.B. 2001, 
in ASP Conf. Ser. 240, 515

\bibitem[Yagi et al.(2002)]{Yagi2002}
Yagi, M., Kashikawa, N., Sekiguchi, M., Doi, M., Yasuda, N., 
Shimasaku, K., and Okamura, S., 2002, \aj, 123, 66

\bibitem[Yagi et al.(2010)]{Yagi2010}
Yagi, M., et al. 2010, \aj, 140, 1814 

\bibitem[Zacharias et al.(2005)]{Zacharias2005}
Zacharias, N., Monet, D.G., Levine, S.E., Urban, S.E., Gaume, R., 
and Wycoff, G.L. 2005, VizieR Online Data Catalog, 1297, 0

\end{thebibliography}
\end{document}